\documentclass[12pt]{article}
\usepackage[usenames,dvips]{pstcol}
\usepackage{color}
\usepackage{amssymb}
\usepackage{epsfig}
\usepackage{footnote}
\usepackage{longtable}
\usepackage{subfigure}
\usepackage{verbatim}
\usepackage{graphicx}
\usepackage{epstopdf}

\usepackage{setspace}
\usepackage{lineno}

\def\18F{^{18}F}
\def\epem{$e^+e^-$}

\def\detector_claim_1{Claim 1 of Section~\ref{claims_detector}}
\newrgbcolor{maroon}{.45 0.1 0.1}
\begin{document}

\pagestyle{plain}
%
%
\begin{flushright}
Version v3c (arXiv pre-print) \\
\today
\end{flushright}

%
\begin{center}
{\Large\bf Methods for Simulating TOF-PET in TOPAS Using a Low-Z Medium }\\

\vspace*{0.25in}

Kepler Domurat-Sousa, Cameron Poe\\
{\it Enrico Fermi Institute, University of Chicago}\\
\vspace*{0.05in}

\vskip 0.1in
 {\it  Published in Nuclear Instruments and Methods, Section A}
\end{center}

\vskip-0.5in
%
\setcounter{secnumdepth}{5} 
\setcounter{tocdepth}{5}
%
%

\begin{abstract}
We have used the TOPAS Geant4-based package to create a parametric simulation of a new TOF-PET scanner design based on a low atomic number scintillator. TOPAS was extended to provide complete information on individual particle position, energy, and identity, allowing access to all particle interactions in the simulation. The data were then used as the `truth' information of particle kinematics for the simulation of the TOF-PET scanner, as described in a separate paper. The Derenzo and XCAT phantoms were added to the simulation with in-patient scattering, positron emission placement, and positron diffusion. Running Geant4 simulations through TOPAS requires substantially smaller amounts of user training than would be needed in raw Geant4. 

This paper describes the methods used to simulate the low-Z whole-body scanner at reduced dose. Specifically, this paper covers the simulation of the FDG decay energy spectrum; the simulation of tissue densities, atomic compositions, and radiotracer doses within imaging phantoms with doses; the custom TOPAS extension used to record particle interaction data; computational performance; and required Geant4 libraries used in TOPAS.
\end{abstract}
\newpage

\section{Introduction}
We present simulations using the TOPAS Geant4-based framework~\cite{FADDEGON2020114} of a whole-body TOF-PET detector based on a low-Z scintillator similar to the scanner of Ref.~\cite{SHIDA2021165801}. TOPAS was chosen as the Monte Carlo physics simulation because it is much more user-friendly~\cite{TOPAS_user_support} and transparent than raw Geant4. Among other advantages, the framework supports traceable code for the underlying physics of low-energy (keV) positrons and electrons in matter~\cite{Penelope_and_Option4} and provides the structure to add user analysis code to access the detailed evolution of all particle trajectories.  TOPAS also allows simulations of a highly sophisticated human phantom through its interface to the XCAT package~\cite{xcat_2010_paper}. This paper covers the techniques developed to create simulations of low-Z PET scanner designs, but does not present results. These methods make simulation of a wide variety of other detector designs feasible in TOPAS.

In a low atomic number scintillation medium such as an organic
scintillator~\cite{guo2019status}, the Compton scattering of a 511 keV
gamma ray dominates absorption via the photo-electric effect by a factor of
$10^4$~\cite{cross_sections, Groom2000}, resulting in a chain of scatterings at
successively lower gamma ray energies, with every scattering producing a recoil
electron in the detector medium~\cite{SHIDA2021165801}. Measurement of the locations of the
scatters, the relative angles between successive scatters, the plane of
the scattering, and the deposited energies and directions of the recoil
electrons allows use of the kinematic constraints of the 2-body
Compton scattering process to perform a statistical time-ordering of
the gamma ray interactions. These methods were developed to simulate a scanner using a switchable fluorescent dye or ``switchillator'' such as proposed in Ref. \cite{SHIDA2021165801}. Such a detection method would produce fluorescent tracks where ionizing radiation (such as the Compton recoil electrons) interact with the detector media.

\begin{figure}[t]
	\centering
	\includegraphics[angle=0,width=0.50\textwidth]{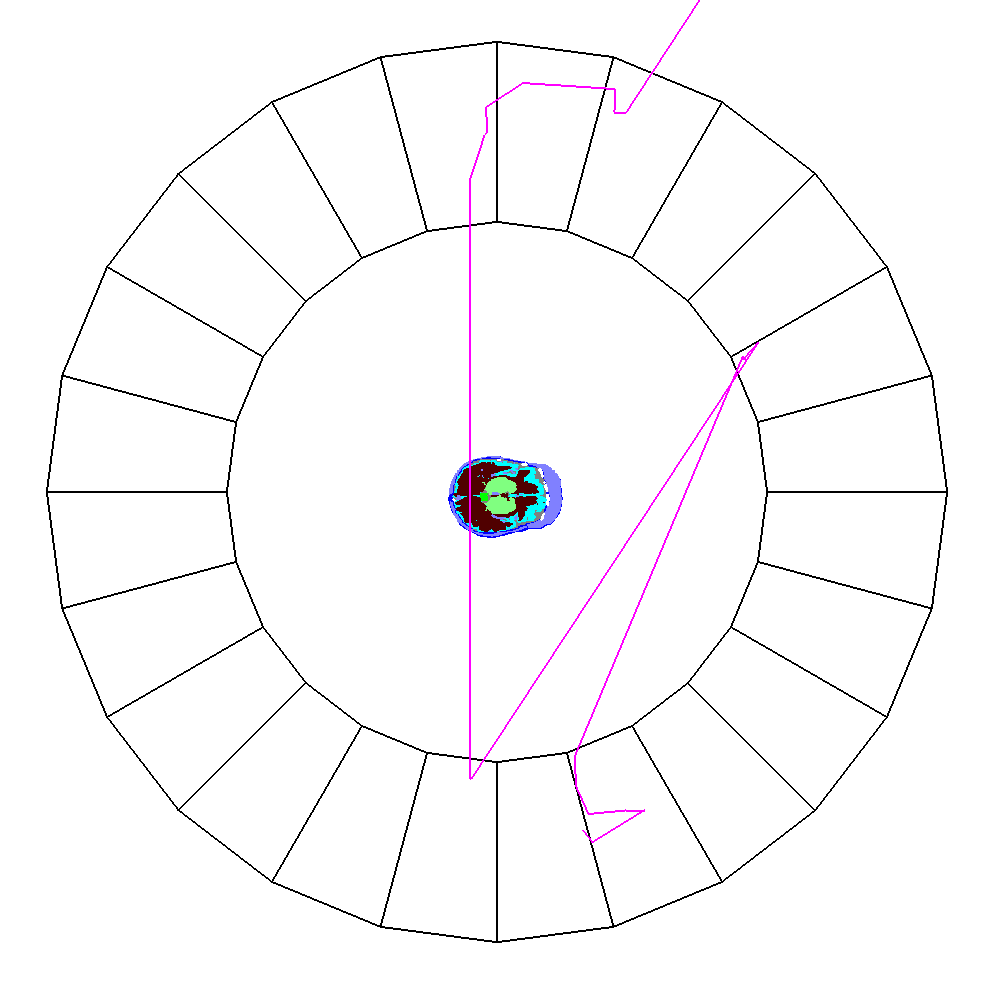}
	\caption{A TOPAS representation of a TOF-PET detector with a
	  superposed simulated \epem annihilation event in an XCAT phantom of a human brain. One gamma ray back-scatters twice across the bore of the detector, and the other eventually exits out the back of the detector after many Compton scatterings. The Line-of-Response (LOR) between the first interactions of the two primary gamma rays is the pink line through the brain. The TOPAS simulation includes the effects of positron range and residual momentum at the time of annihilation on the colinearity of the two gamma rays. Precise time measurements of optical photons would provide additional constraints
on the often complex trajectories of the gamma rays.}
	\label{fig:Detector_and_Brain}
\end{figure}

Figure~\ref{fig:Detector_and_Brain} shows the end view of the TOF-PET detector surrounding a phantom extracted from the TOPAS simulation~\cite{FADDEGON2020114}. The image on the axis is a human brain extracted from the XCAT human whole-body phantom~\cite{xcat_2010_paper}. The trajectories of the two gamma rays from an \epem annihilation in the brain are superposed on the detector. In this event neither of the two gamma rays underwent in-patient scattering; however one gamma ray back-scatters twice across the bore of the detector, and the other eventually exits out the back of the detector after many Compton scatterings. For each scattering in a detector module the time, position, energy, and a voxelized image of the recoil electron track are recorded. Determining the Line-of-Response (LOR) between the first interactions of the two primary gamma rays is the reconstruction problem to be solved~\cite{domuratsousa2023simulation}.

\section{The TOPAS Geant4-based Simulation Framework}
\label{TOPAS}
\begin{figure}[ht]
    \centering
    \includegraphics[angle=0,width=0.3\textwidth]{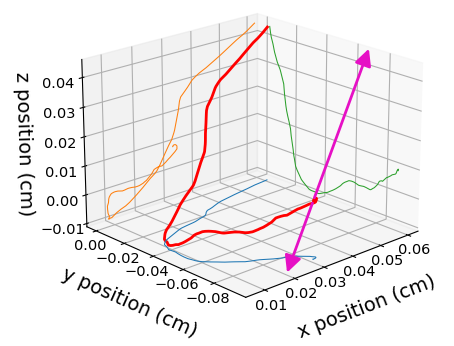}
    \hfil
    \includegraphics[angle=0,width=0.3\textwidth]{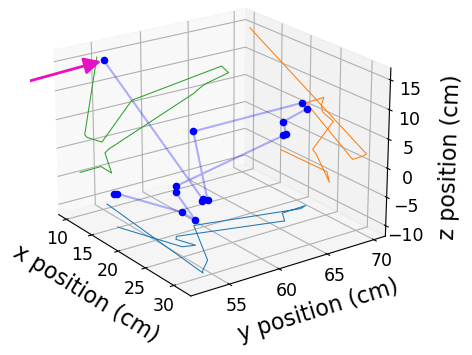}
    \hfil
    \includegraphics[angle=0,width=0.3\textwidth]{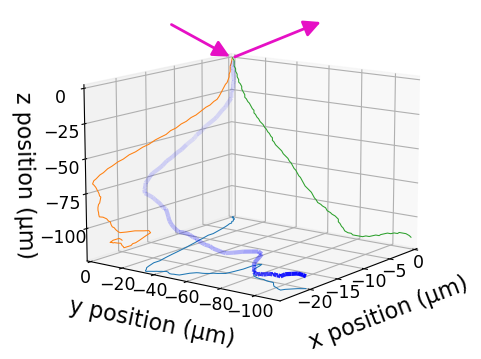}
    \caption{Left: The positron path from creation to annihilation. The gammas from the positron annihilation are shown in pink. Center: A chain of successive Compton scatterings from a 511 keV gamma ray (arrow). Right: The electron from a Compton scattering in the simulated detector. The gamma path is shown in pink and the electron is shown in blue. The visual density of the electron path shows the amount of energy deposited by the electron into the module at each step.}
    \label{fig:Chain_of_Scatters_One_Module}
\end{figure}

TOPAS~\cite{FADDEGON2020114} is a user interface for applications in medical physics to the CERN-based Geant4~\cite{Geant4} toolkit for the simulation of the passage of particles through matter. It allows configuring Geant4 using text parameter files defining detector and patient geometries, materials, particle sources, and the organization of the simulated data.
The TOPAS framework allows one to access all required simulation capabilities of Geant4 without having to master the full low-level Geant4 code.

The user-friendly interface of TOPAS allows the use of complex geometries while providing several protections against all-too-common Geant4 user errors. TOPAS checks for overlaps between geometric volumes in the simulated objects, forestalling  Geant4 crashes. TOPAS also provides strict typing for values given in parameter definitions
so that, for example, a distance in cm cannot be entered as a value for an angle parameter. Because TOPAS is well-documented, robust, and provides excellent user support~\cite{TOPAS_user_support}, new features can be rapidly incorporated and fully tested.

TOPAS has enabled us to build a full simulation of a parameterized whole-body PET scanner with changeable phantoms. During detector initialization, the relevant Geant4 physics modules, the maximum step sizes in the detector volume, the selection criteria, the lists of parameters and variables to be recorded for analysis, and the output data format are all defined. In total, defining the detector in its current form takes less than a hundred lines of easily human-readable information~\cite{Domurat-Sousa_TOPAS_Simulation_of_2023}.

Figure~\ref{fig:Chain_of_Scatters_One_Module} shows the TOPAS simulation of the Compton scatter chain, an electron from one scatter traveling through the detector medium (Section \ref{TOPAS_Detector}), and the positron diffusion in the phantom (Section \ref{txt:positron}).

\subsection{Physics Packages}
The simulation was run in TOPAS 3.8 with the underlying physics of low-energy (keV) electrons and gamma rays in matter determined in two Geant4 physics modules:  ``g4em-standard\textunderscore opt4" and  ``g4em-penelope" ~\cite{Penelope_and_Option4}.

\subsection{Detector Geometry and Medium}
\label{TOPAS_Detector}
An example low-Z whole-body PET scanner with a simulated annihilation event from TOPAS superimposed is shown in Figure~\ref{fig:Detector_and_Brain}. The location of the positron annihilation is inside the brain; the successive interactions of the gamma in the detector by Compton scattering are found by the resulting Compton electron tracks in the low-Z medium, as shown in Figure~\ref{fig:Chain_of_Scatters_One_Module}. The location of the first scatter of each gamma ray determines the corresponding end of the Line-of-Response (LOR)~\cite{SHIDA2021165801}.

The cylindrical detector has a bore radius of 45 cm\footnote{The large bore radius is to accommodate the voxelized XCAT phantom, which otherwise intersects the detector.} and a length of 2 m. The detector active volume is 30 cm radially. The detector material was modeled as Linear Alkyl-Benzine (LAB)~\cite{LAB}. Components present in trace quantities such as a switchillator were not modeled, as they would have minor effects on the interaction cross-sections in the overall detector medium~\cite{SHIDA2021165801}.

The behavior of many common materials are defined in the Geant4 book for application developers ~\cite{Geant4_Materials}, and it is possible to use these definitions directly in TOPAS. For materials that are not predefined the user can create them using the parameter setup in a similar manner to the rest of the simulation. LAB was defined as a material in TOPAS with mass fraction 87.9\% Carbon, 12.1\% Hydrogen. The mean excitation energy for LAB was estimated to be 59.4 eV, based on the chemical structure.

The TOPAS graphics system was used to check that the geometry is as expected, as shown in Figure \ref{fig:Detector_and_Brain}.

\subsection{Modeling the Positron Energy Spectrum}
\label{txt:positron}
\begin{figure}[ht]
    \centering
    \includegraphics[angle=0,width=0.55\textwidth]{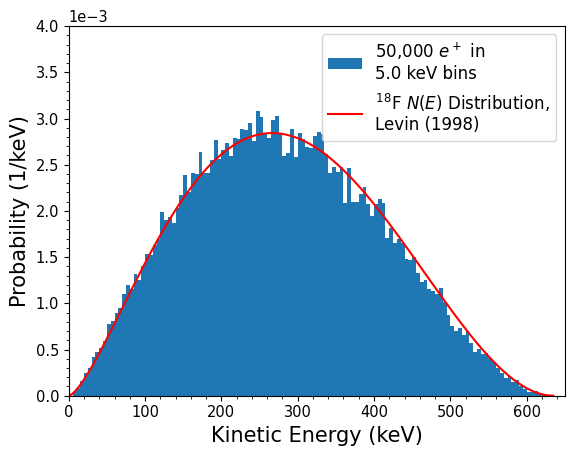}
    \caption{Probability distribution of the positron energy spectrum at decay for Fluorine-18. The red curve is the predicted spectrum following Equation \ref{eq:cs-wu-eq}; the blue histogram is from data taken from the TOPAS simulation.}
    \label{fig:energy-spectrum-positron}
\end{figure}

The radiotracer used in the simulations was fluorodeoxyglucose (FDG). To model this radiotracer accurately in the phantoms, it was necessary to be able to both change the concentration of FDG in various tissues or materials and recreate the positron energy spectrum at decay. TOPAS has native support for both of these challenges.

TOPAS' volumetric particle source was used to simulate FDG decays in the phantoms. The volumetric source is a category of particle source that creates particles at randomly sampled positions throughout a volume, with the user specifying the material to be sampled within the volume, the particle to be created at those positions, and the energy distribution of those particles. The volumetric source also lets a user specify the total number of particles to create within the volume. Instead of setting the activity of a tissue to a rate density (FDG decays per unit time per unit volume), a time interval was chosen over which the simulated scan would take place, and the decay rate was averaged over that period to get a number of FDG decays per unit volume. This procedure specifies different activities in different tissues since, for example white matter and gray matter in the XCAT phantom brain have different uptake ratios of FDG.

The positron energy spectrum was created using the discrete energy spectrum type with variable energy. TOPAS can simulate any arbitrary energy spectrum by letting users assign weights to discrete energy values. Following Levin and Hoffman \cite{Levin1999-pk}, the energy spectrum function for Fluorine-18 was modeled as:
\begin{equation}
    \label{eq:cs-wu-eq}
    N(E)dE = C(E_{\rm{max}} - E)^2 E^2 \left(1 - exp\left(2 \pi Z \alpha \sqrt{\frac{E}{2m_e}} \right) \right)dE
\end{equation}
where $E$ is the energy of the positron, $C$ is a normalization constant to satisfy $\int_0^{E_{\rm{max}}} N(E)dE = 1$, $E_{\rm{max}} = 635$ keV is the maximum energy a positron can have in the decay \cite{max_positron_energy}, $Z=9$ is the atomic number of Fluorine, $\alpha$ is the fine-structure constant, and $m_e$ is the mass of the positron. Since the maximum positron decay energy is less than 635 keV, the energy spectrum was set in intervals of 1 keV from 0 to 635 keV and set different weights to each interval corresponding to the probability density function in Equation \ref{eq:cs-wu-eq}. Figure~\ref{fig:energy-spectrum-positron} shows a comparison of the exact energy spectrum equation and data taken from TOPAS of simulated positron energy at decay.

\subsection{Custom TOPAS Scorer}
TOPAS does not directly provide the information on the location and energy of the electron paths for the Compton reconstruction. The problem stems from TOPAS' origin as a tool for proton therapy and similar radiation exposure situations, where the object is to calculate the total dose information, rather than the
interactions, ionization, and trajectories of individual particles. Fortunately, TOPAS does provide a very complete extension system that allows building a tuple containing detailed information on the electron and gamma ray trajectories and interactions in each event. The tuple dataset is recorded for later analysis by dedicated code to characterize the detector performance and to produce plots such as the positron behavior in the active medium shown in Figure~\ref{fig:energy-spectrum-positron} or the Compton chain in Figure \ref{fig:Chain_of_Scatters_One_Module}.

The custom tuple provides the following information: the history number; the particle energy at the beginning of the step; the energy the particle deposits in the step; the x, y, z, and time of the step recorded at the beginning of the step; the particle type (in PDG format); and an internal particle count to distinguish particles of the same type.

\begin{figure}[t]
	\centering
	\includegraphics[width=0.49\textwidth]{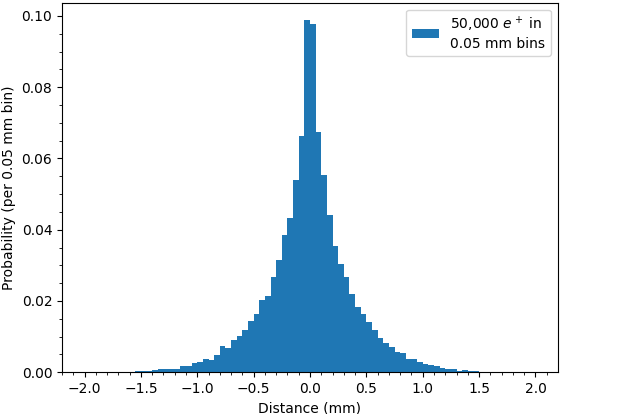}
	\hfil
	\includegraphics[width=0.49\textwidth]{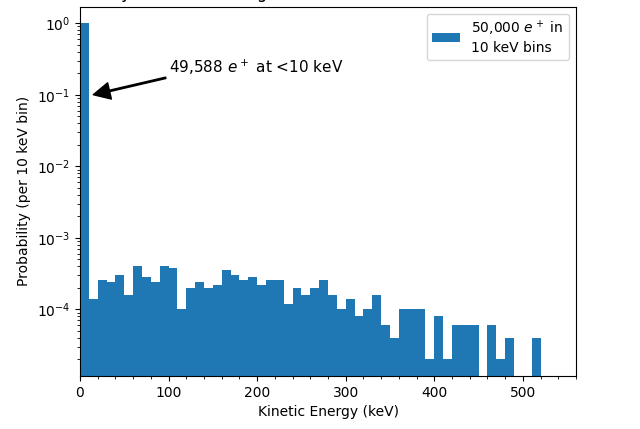}
	\caption{Simulated data from TOPAS showing the 1D displacement and energy of positrons at annihilation. Both histograms are from runs with 50,000 positrons. Left: The distance traveled along the x-axis by positrons from creation to annihilation. Right: A log plot of the energy at annihilation of the positrons. The vast majority of positrons annihilate with no residual energy. In the lowest bin, 49584 of 49588 events have identically zero energy.}
	\label{fig:positron_position_and_final_energy}
\end{figure}

\subsection{Functional Description of the Simulation}
\label{functional_description}
For simulation runs where the goal is a reconstructed image, only the data that would be measured by the detector was recorded. To make bookkeeping simpler, both the gamma ray interactions and the ionized particle interactions were stored. For runs exploring the performance of the Compton reconstruction system, the software collects additional `truth' information. This information includes all particle interactions in the simulation, allowing tracking of in-patient scatters and the positron trajectory and momentum.\footnote{The voluminous truth information is written out only on diagnostic runs with fewer events.} Figure~\ref{fig:positron_position_and_final_energy} shows distributions in the displacement and energy of positrons at annihilation\cite{Angular_distribution_DeBenedetti} as generated by TOPAS using Penelope and the appropriate Geant4 packages~\cite{Penelope_and_Option4}. The distributions agree well with published results~\cite{Levin1999-pk, Positron_multiple_scattering_Blanco}.\footnote{This is largely a check for gross operator errors, as there is most likely a very high degree of correlation between the codes.}

\subsection{The Derenzo and XCAT Brain Phantoms}
\label{phantoms}
Two phantoms were simulated: the Derenzo geometric phantom~\cite{Derenzo} made of active rods in a surrounding background volume, and the XCAT human phantom \cite{xcat_2010_paper}. A 20-mm diameter tumor was inserted into the XCAT brain phantom~\cite{brain_tumor_size}.  Both phantoms were centered on-axis in the detector.

If multiple volumetric sources in TOPAS have histories in a single run then multiple positrons with identical identification numbers will be created in a history, leading to a pileup in the output file. To simplify bookkeeping, sources were separated into different TOPAS runs using time features such that a source is only on for a specific internal time. Each source is given a different time and then the internal TOPAS simulation clock is stepped across all of the needed times. While this procedure can create multiple histories of the same number, they are separated in the file by histories with a different value.

\begin{figure}[!ht]
\centering
\includegraphics[angle=0,width=0.32\textwidth]{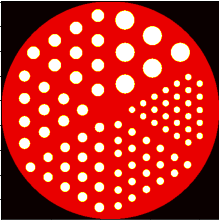}
\hfil
\includegraphics[angle=0,width=0.32\textwidth]{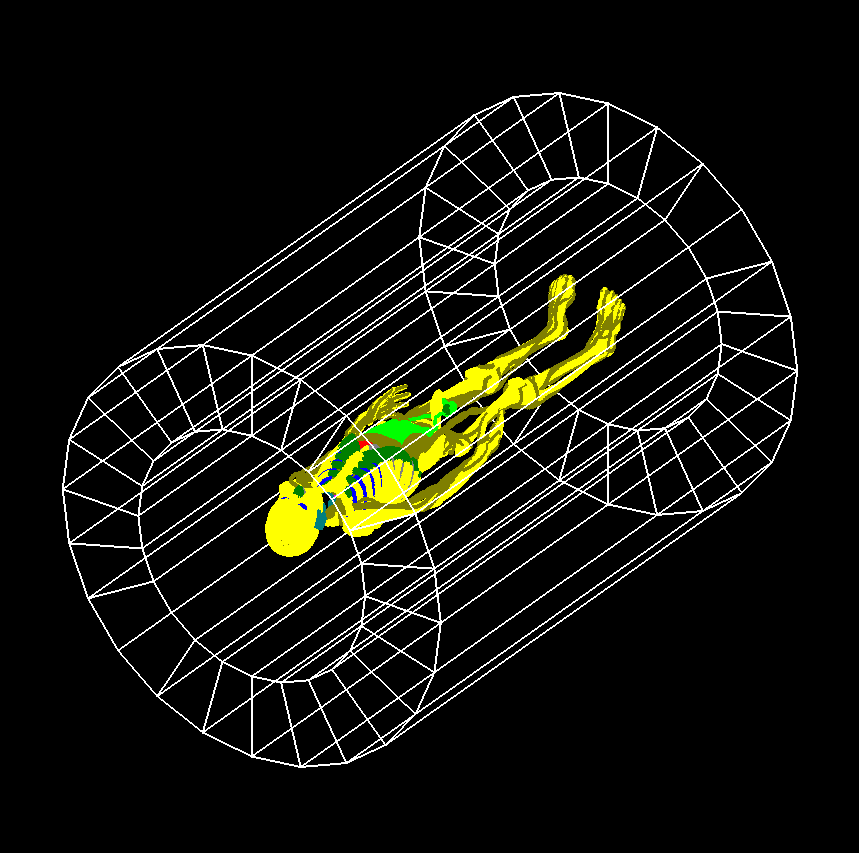}
\hfil
\includegraphics[angle=0,origin=c,width=0.32\textwidth]{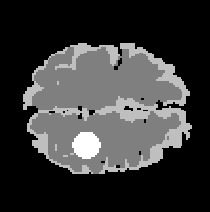}

\caption{The input images of the phantom geometries placed in the PET scanner, viewed in the TOPAS graphical interface. Left: Transverse slice showing the distribution of rods in the Derenzo phantom. Center: The full XCAT whole-body phantom in the cylindrical detector. The radius inner radius, outer radius, and length of the detector are 0.45 m, 0.75 m, and 2 m, respectively. Right: Transverse slice of the XCAT brain including a 20 mm diameter tumor.}
\label{fig:Derenzo_XCAT_truth_images}
\end{figure}

Figure~\ref{fig:Derenzo_XCAT_truth_images}
shows the input images of the phantom geometries as placed in the PET scanner, viewed in the TOPAS graphical interface.

\subsubsection{Implementing the Derenzo Phantom}
The Derenzo Phantom was created using only the TOPAS native cylinder object. A Python program was written that creates a single large cylinder as the main body of the phantom. The code then creates additional smaller cylinders to act as the high activity rods in the phantom. These rods are defined based on a file containing location and radius information of each rod within the phantom. All of the rods are defined as children of the main body. This allows specifying only the position of the main body of the Derenzo phantom in the detector. All of the objects are made of water, using the ``G4\textunderscore WATER'' material. Each cylinder is assigned a volumetric source, allowing us to change the number of positrons created in each geometry component. To give a simple and easily modifiable activity behavior, a variable number of positrons/volume is defined separately for the rods and the background, and this value is multiplied by the volume of the geometry component to give total positrons. Each rod and the background is given a separate internal time to avoid readout conflicts.

\subsubsection{Implementing the XCAT Phantom}
\label{xcat_desc}
The 4D Extended Cardiac-Torso (XCAT) Phantom Version 2.0 is a software program that can simulate detailed patient anatomies for use in medical imaging~\cite{xcat_2010_paper}. XCAT outputs voxelized male and female phantoms representative of a 50th percentile U.S. adult. Further anatomical variation and customization options are available. In addition, XCAT can simulate cardiac plaque, cardiac defects, and arbitrary spherical lesions~\cite{xcat_2010_paper}.

Simulating XCAT phantoms is straightforward in TOPAS. Users supply the XCAT software with a parameter file that assigns a value of the tracer activity associated with each tissue. XCAT generates a binary output file corresponding to a voxelized phantom where each voxel is assigned the appropriate activity. At run time TOPAS inputs the voxelized XCAT phantom file in parallel with the list relating the numeric activities to tissue names and a second list relating tissue names to material properties (e.g. atomic composition, mass density). TOPAS then renders the voxelized phantom with proper positron densities in the Geant4 simulation according to Section \ref{txt:positron}

\subsection{Computational Performance}
\label{computation}
As is customary for data with high-resolution-image-producing detectors~\cite{Pion_discovery}, the data acquisition, in this case the TOPAS generation of simulated data, was separated from the subsequent analysis.  The simulation  writes files containing the run parameters and lists of annihilation events (aka `list mode').
A typical simulation of a given geometry and activation comprises 124 million positrons for the Derenzo format at an
activation of 0.01 of standard PET doses~\cite{domuratsousa2023simulation}, and has a total data size
of 405 GBytes. The corresponding numbers for the XCAT brain, also at 1\% dose, are 83 million positrons and a total data size of 157 GBytes.

A TOPAS simulation of 83 million events on the Derenzo phantom takes 15 hours using 14 threads on an Intel Xeon CPU E5-2620 v4 @ 2.1 GHz. The XCAT brain phantom simulation in TOPAS is currently appreciably slower, requiring 68 hours for the same number of events. The computational load is likely due to the voxelization in the current implementation of the XCAT phantom.

\section{Summary}
\label{Summary}
We have taken advantage of the TOPAS simulation framework to write a parametric simulation of a TOF-PET detector based on an ionization-sensitive low-Z persistent medium and fast MCP-based photodetectors to record the chain of Compton scatterings of the annihilation gamma rays.  The TOPAS framework enables building a full simulation of a parameterized whole-body PET scanner with changeable phantoms, and a straight-forward selection of Geant4 physics modules and the initialization of maximum step sizes in the detector volume, the selection criteria, the lists of parameters and variables to be recorded for analysis, and the output data format.

TOPAS interfaces natively with the XCAT phantom package, and also contains tools for creating geometric phantoms.
Two phantoms were simulated: the XCAT human brain, and the Derenzo geometric phantom.
The Geant4-based simulation includes the low-energy range and residual momentum of the positron, in-patient scattering,
Compton scattering of the gamma-rays in the detector, and tracking of the ionization deposited in the active
medium by the Compton electrons.

The combination of ease of use and flexibility that TOPAS provides allows for novel detector concepts to be quickly simulated. The above techniques have been used for simulations of a range of detectors and makes rapid iteration on detector concepts possible. Finally, the extremely usable nature of TOPAS allows undergraduate students with a minimum of training to simulate complex detectors.

\clearpage

\section{Acknowledgements}
For their exemplary software development and user support, we thank both Joseph Perl (TOPAS)
and Paul Segars (XCAT). We thank Mary Heintz for essential computational system support. We thank Henry Frisch, Allison Squires, and Patrick La Riviere for their exceptional guidance through this work.

K. Domurat-Sousa, and C. Poe were supported by the
University of Chicago College, Physical Sciences Division, and Enrico Fermi Institute, for which we thank Steven Balla and Nichole Fazio, Michael Grosse, and Scott Wakely, respectively. C. Poe was additionally supported by the Quad Undergraduate Research Scholars program.

\enlargethispage{6.0in}
%

\newpage
\clearpage


\bibliographystyle{elsarticle-num}
\bibliography{low-z_methods.bib}

\end{document}